\documentclass[conference]{IEEEtran}
\IEEEoverridecommandlockouts
\usepackage{cite}
\usepackage{amsmath,amssymb,amsfonts}
\usepackage{multicol}
\usepackage{supertabular,booktabs}
\usepackage{graphicx}
\usepackage{textcomp}
\usepackage{xcolor}
\usepackage{url}
\usepackage{tikz}
\tikzstyle{vertex}=[circle, draw, inner sep=0pt, minimum size=6pt]
\newcommand{\vertex}{\node[vertex]}
\usetikzlibrary{arrows,decorations.markings}
\usetikzlibrary{shapes}
\usetikzlibrary{calc}
\usetikzlibrary{patterns}
\usepackage{algorithm}
\usepackage{algorithmicx}
\usepackage{algpseudocode}
\usepackage{longtable}
\def\BibTeX{{\rm B\kern-.05em{\sc i\kern-.025em b}\kern-.08em
		T\kern-.1667em\lower.7ex\hbox{E}\kern-.125emX}}
\begin{document}
\title{An Application of Random Walk on Fake Account Detection Problem: A Hybrid Approach}

\author{\IEEEauthorblockN{Ngoc C. L\^{e}}
	\IEEEauthorblockA{\textit{School of Applied Mathematics and Informatics} \\
		\textit{Hanoi University of Science and Technology} \\
		\textit{Institution of Mathematics} \\
		\textit{Vietnam Academy of Science and Technology} \\
		lechingoc@yahoo.com}
	\and
	\IEEEauthorblockN{Manh-Tuan Dao}
	\IEEEauthorblockA{\textit{iCOMM Media \(\& \) Tech, Jsc}\\
		tuan.dao@icomm.vn}
	\and 	
	\IEEEauthorblockN{Hoang-Linh Nguyen}
	\IEEEauthorblockA{\textit{School of Applied Mathematics and Informatics}\\
		\textit{Hanoi Univ of Science and Technology} \\
		linh.nh5015@gmail.com}
	\and 	
	\IEEEauthorblockN{Tuyet-Nhi Nguyen}
	\IEEEauthorblockA{\textit{School of Applied Mathematics and Informatics}\\
		\textit{Hanoi Univ of Science and Technology} \\
		nguyennhiaktf@gmail.com}
	\and 
	\IEEEauthorblockN{Hue Vu}
	\IEEEauthorblockA{\textit{School of Applied Mathematics and Informatics}\\
		\textit{Hanoi Univ of Science and Technology} \\
		hue.hnue@gmail.com}
}
\maketitle
\begin{abstract}
Social networks play a significant role in today's world. The importance of social networks, for example Facebook or Twitter, are undeniable. However, they also have many issues. One of which is the need for a defense mechanism against fake accounts. It is obviously not a trivial task to separate fake accounts from authentic ones. In this paper, we propose a ranking scheme, comprising of both graph based and feature based approaches to aid the detection of fake Facebook profiles. Utilizing Support Vector Machine (SVM) \cite{cortes1995} and SybilWalk \cite{JWZ17}, the model achieved high accuracy over the set of ten thousands Vietnamese Facebook accounts.
\end{abstract}
\begin{IEEEkeywords}
Fake Account, Network Theory, Ramdom Walk, Support Vector Machine, SVM, SybilWalk
\end{IEEEkeywords}
\section{Introduction}
The last decade witnessed dramatic growth in size as well as influence of online social networks (OSNs) such as Twitter, LinkedIn and especially Facebook. As of 2018, Facebook has more than two billions active users. For better or worse, these sites have had a huge impact not only on social interaction, but also on education, employment, business, etc. Communication and information sharing are easier than ever. However, what follows is a lot of issues with privacy, cyber bullying, social engineering, online impersonation and so on.\\ 
A fake account can be defined as an account which is not representative of a real person or organization. This is not to be confused with clones, whose identity is that of an actual person but possessed by some others for malevolent deeds. Facebook estimated up to six to ten percents of its user base are either fake or duplicate accounts in 2017 \cite{BAH17}. However, this number can greatly fluctuate since there are a lot of new ones being created everyday and Facebook taking measures to cope with them. Presumably, fake accounts are still very much elusive to Facebook security measures, known as Facebook Immune System (FIS) \cite{FIS,ASONAM}. The detection of fake accounts remains a problematic case for Facebook as well as in social network security research.\\
Since false positives can heavily damage the experience of users if actions are taken to suspend accounts assumed non - genuine right away, the task of filtering out fake accounts has not been successfully brought to automation. Social networks providers have had to resort to inefficient and costly manual labor. For example, Tuenti Technologies employs an inspection team which must review well over ten thousands reports per day. However, only about 5\% of the reviewed accounts are indeed fake \cite{Sybil}.\\
In this paper, we present a procedure to help identify fraudulent accounts (human inspections and decision making are still required). We tried to capture both the characteristics of fake profiles as well as the relationships between these and the authentic profiles. The result is promising over a set of twelve millions accounts of the test set.\\
The rest of the paper is organized as following. Section \ref{backgr} introduces the background knowledge and reviews some of the related works. In Section \ref{sec:model}, the details about the proposed model and the features selected for machine learning modules are given. Results are given in section \ref{result}. Section \ref{conclusion} gives some perspectives and comments about effectiveness and limitations of the scheme, as well as future directions.
\section{Background and Related work}\label{backgr}
\subsection{Background}
\subsubsection{Support Vector Machines (SVM)}
Support Vector Machines is a supervised learning method mostly used for classification and regression analysis. Given a training dataset $\{X_i, Y_i\}^n_{i=1}$, where $X_i$ represents the $n$-dimensional input vector and $Y_i\in\{1,-1\}$ represents the class or label memberships (also positive and negative samples), a decision hyperplane is constructed that best separates the two classes in the sense that it has the largest distance to the nearest training-data point of any class (also known as functional margin). SVM can perform both linear as well as non-linear classification, with the latter requires a little more data preprocessing through the so-called kernel function.\\
The decision hyperplane obtained from training is defined by the equation
\[c^\top x - b = 0 \]
where $c\in \mathbb{R}^n, b\in \mathbb{R}$. Given a new input vector $X$, we can, for example, classify $X$ into label $1$ if $c^\top X - b < 0$ and into label $-1$ if $c^\top X - b \geq 0$.
\begin{center}
	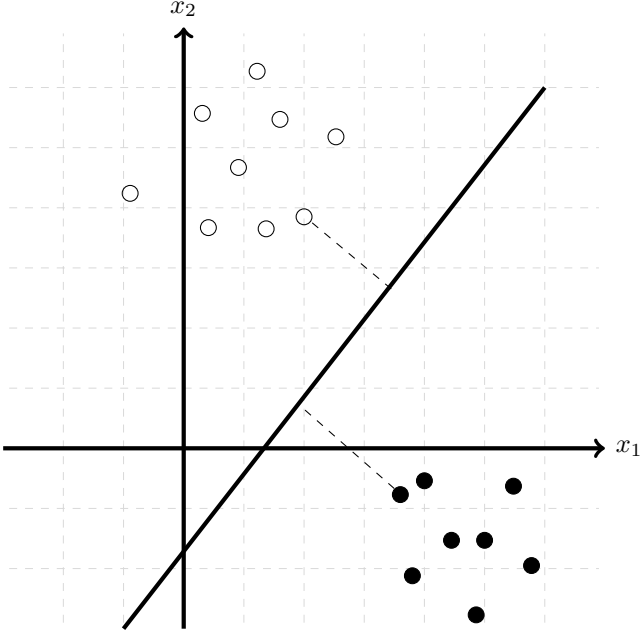
\begin{figure}[h]
		\begin{tikzpicture}[scale = 0.8]
		\draw[help lines, color=gray!30, dashed] (-4.9,-4.9) grid (4.9,4.9);
		\draw[->,ultra thick] (-5,-2)--(5,-2) node[right]{$x_1$};
		\draw[->,ultra thick] (-2,-5)--(-2,5) node[above]{$x_2$};
		\draw[ultra thick] (4,4)--(-3,-5);
		\vertex[] (p1) at (0,1.85) {};
		\vertex[] (p1) at (0.53,3.18) {};
		\vertex[] (p1) at (-0.63,1.65) {};
		\vertex[] (p1) at (-1.09,2.67) {};
		\vertex[] (p1) at (-0.78,4.27) {};
		\vertex[] (p1) at (-2.89,2.24) {};
		\vertex[] (p1) at (-1.69,3.57) {};
		\vertex[] (p1) at (-0.4,3.47) {};
		\vertex[] (p1) at (-1.59,1.67) {};
		\draw[dashed] (0.135,1.75) -- (1.45,0.65);
		\draw[dashed] (1.5,-2.67) -- (0,-1.35);
		\vertex[fill = black] (p2) at (1.6, -2.77) {};
		\vertex[fill = black] (p2) at (2, -2.54) {};
		\vertex[fill = black] (p2) at (1.8, -4.12) {};
		\vertex[fill = black] (p2) at (2.86, -4.77) {};
		\vertex[fill = black] (p2) at (3, -3.53) {};
		\vertex[fill = black] (p2) at (3.48, -2.63) {};
		\vertex[fill = black] (p2) at (2.45, -3.53) {};
		\vertex[fill = black] (p2) at (3.78, -3.95) {};
		\end{tikzpicture}
		\caption{\em Example of a linear Support Vector Machine}
	\end{figure}
\end{center}
Let $z = c^\top X - b$, we can normalize $z$ with the sigmoid function
\[S(z) = \frac{1}{1 + e^{-z}}.\]
As $z$ approaches positive infinity, $S(z)$ approaches 1. In the case $z$ approaches negative infinity, $S(z)$ approaches 0. We can let $S(z)$ represents the possibility that $z$ has label $-1$, which is exactly what we are going to do in our detection scheme.	
\subsubsection{Social network graph}
A graph is a discrete structure used to model objects and pairwise relations between them. 
Graph theory is a strong tool when it comes to networks researches. A graph, $G = (V, E)$, is a pair of vertex set $V$ edge set $E$.\\
For our application, we shall model the social network Facebook as a graph, in which each node $u\in V$ represents a user and each edge $e\in E$ represents an established relationship (e.g friendship, commenting in the same posts, ...) between two nodes. We may also denote an edge by its two end nodes, i.e $uv$. The graph is undirected and contains no loop (an edge connecting a node with itself). The degree of node $u$ is the number of edges connected to $u$. $adj(u)$ is the set of nodes adjacent to $u$ (connected to $u$ by an edge).\\
\begin{figure}[h]
	\begin{center}
		\begin{tikzpicture}
		
		\vertex[](p1) at (0.3, 0.3) {};
		\vertex[](p2) at (-0.2, 0.5) {};
		\vertex[](p3) at (-0.5, 0.8) {};
		\vertex[](p4) at (-0.4,-0.5) {};
		\vertex[](p5) at (-1.5,0.3) {};
		
		\vertex[](p6) at (-1,-2) {};
		\vertex[](p7) at (-2.6,-4) {};
		\vertex[](p8) at (1,-3) {};
		\vertex[](p9) at (1,-2) {};
		\vertex[](p10) at (-1.4,-3.5) {};
		\vertex[](p11) at (-1.45,-2.5) {};
		\vertex[](p12) at (0,-3) {};
		\vertex[](p13) at (0.5,-3.5) {};
		
		\draw [] (p1) to (p2); \draw [] (p1) to (p4); \draw [] (p1) to (p5); \draw [] (p4) to (p6); \draw [] (p4) to (p3); \draw [] (p4) to (p5);
		\draw [] (p3) to (p6);
		\draw [] (p2) to (p4);\draw [] (p2) to (p3); \draw [] (p5) to (p3);
		\draw [] (p6) to (p10);\draw [] (p11) to (p10); \draw [] (p11) to (p6); \draw [] (p9) to (p6); \draw [] (p7) to (p11);
		\draw [] (p9) to (p13);\draw [] (p9) to (p10);\draw [] (p9) to (p12);
		\draw [] (p10) to (p12);\draw [] (p9) to (p11); \draw [] (p12) to (p13); \draw [] (p8) to (p12); \draw [] (p8) to (p9); \draw [] (p7) to (p10);
		
		\vertex[fill = black](p14) at (5,0.5) {};
		\vertex[fill = black](p15) at (5.5,0) {};
		\vertex[fill = black](p16) at (4.3,-0.1) {};
		\vertex[fill = black](p17) at (4.5,-0.8) {};
		
		\draw [] (p14) to (p15); \draw [] (p16) to (p17);  \draw [] (p14) to (p17);
		\draw [] (p16) to (p15); \draw [] (p16) to (p1);
		
		\vertex[fill = black](p18) at (5.1,-3) {};
		\vertex[fill = black](p19) at (4,-3.5) {};
		\vertex[fill = black](p20) at (4.3,-3.8) {};
		\vertex[fill = black](p21) at (4.1,-2.7) {};
		
		\draw [] (p18) to (p17); \draw [] (p18) to (p20);\draw [] (p18) to (p21); \draw [] (p19) to (p21); \draw [] (p19) to (p20); \draw [] (p19) to (p15);
		\draw [] (p19) to (p9);
		
		\node[draw] at (-0.7, -5){Real accounts};
		\node[draw] at (4.8, -5){Fake accounts};
		
		\node[draw, ellipse] at (2.5, 0.7) {Attack Edge};
		
		\end{tikzpicture}
	\end{center}
	\caption{\textit{The social network graph}}
\end{figure}
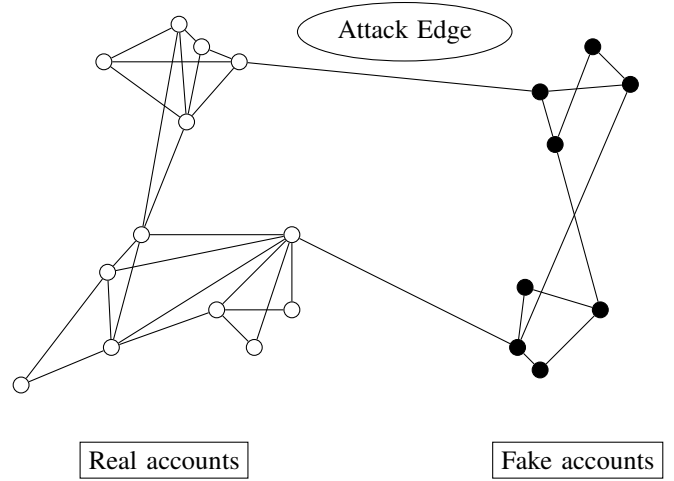
The term \textit{Sybil} which is used widely in graph-based network security research refers to a forged, pseudonymous identity in peer-to-peer networks. It was derived from the book of the same name, a case study of a woman diagnosed with dissociative identity disorder \cite{schreiber1973}. Later researches into social networks impersonation also use this term to refer to a fake node in the users network graph. An \textit{attack edge} is an edge connecting a benign user and a Sybil.
\subsubsection{Random Walk}
A random walk \cite{pearson1905} on a graph $G$ can be described as a succession of nodes $u_1, u_2, ..., u_k$, where each node is chosen from the neighbors of the last randomly. Given that a random walk is long enough, it can land on any node of the graph with uniform degree-normalized probability (the probability that the random walk ends up in $u$ divided by the degree of $u$ is roughly the same for every $u$). This is known as the convergence of a random walk to its stationary distribution. A graph is said to have \textit{fast-mixing} property if this convergence happens in a relatively small number of steps. In our application we shall use a variation of a random walk algorithm developed in \cite{JWZ17}.
\subsection{Related work}
So far, the approaches for the Sybil accounts detection problem can be divided in to two groups: (1) Feature-based approaches using features of accounts and (2) Graph-based approaches using the relations between accounts.
\subsubsection{Feature-based approaches}
Feature-based (e.g Machine-Learning based) methods have long been used in OSNs security. Take spam detection for example, in \cite{junkmail}, the authors first proposed a Bayesian approach to filter spam emails considering domain-specific features. Since then, spam mail filtering techniques has matured over time and achieved high accuracy. However, it is an entirely different challenge when moving from the context of email systems to massive networks. As evident in \cite{automatedSybil}, automated Machine-Learning-based fake account detection suffers from high false negative and positive rates. Much similar to why Machine Learning has not been effective in network intrusion \cite{intrusion}, these approaches could not fully cover the diverse activities and properties of intruders, and are subject to overfitting. High false positive rate is particularly harmful to social networks providers, as users definitely do not respond well to their account being wrongly suspended. Besides, there are also issues with scalability, lacking in flexibility (attackers can easily adapt to avoid traits recognized by the classifiers),... just to name a few reasons why there has not been a feasible solution.
\subsubsection{Graph-based approaches}
Graph-based Sybil detection has long been studied in peer-to-peer systems. As stated every networks can be modeled as a graph $G = (V,E)$. Graph-based solutions, also called random walk - based solutions, rely on social graph properties to uncover fake users. Notable examples include \cite{GP,SybilLim,SybilGuard}.\\
Presumably, Sybils have a disproportionately small number of connections to real users. Existing works are largely based on this assumption \cite{Sybil,GP}. Naturally, graph-based solutions uncover Sybils from the perspectives of known non-Sybil nodes. Take SybilInfer\cite{GP} for example, a set of traces $T$ are generated and stored by performing special random walks over the social graph $G$. Once the probabilistic model is defined, calculate for any set of nodes $X$ and the generated trace $T$, the probability that $X$ consists of honest nodes. We can calculate the probability of any node in the system being honest or dishonest. SybilGuard \cite{SybilGuard} and SybilLimit \cite{SybilLim} also infer Sybils based on a large number of random walk traces. SybilRank \cite{Sybil}, widely used in many applications, outputs perceived likelihood of a node being fake. It relies on the observation that an \textit{early-terminated} random walk starting from a non-Sybil node has a higher degree normalized probability to land at a non-Sybil node than a Sybil node. From a collection of know benign users, known as trust seeds, SybilRank then uses short random walks to assign trust score to other nodes. Unfortunately, the problem of multi-community structure in social graphs (high connectivity in each community but low inter-community connectivity), imposed difficulty as non-Sybils that do not belong to the communities of trust seeds may be mistaken for Sybils.\\
Given that a graph has fast-mixing property and \textit{homophily} property \cite{levin2017} (two nodes sharing a same edge has high probability of belonging to the same class), graph-based methods have guaranteed performance and accuracy \cite{JWZ17}. However, these assumptions do not always hold in the case of real world social graph. Leveraging only either benign users or Sybils also limits the potential of these methods. 
\section{Proposed model}\label{sec:model}
The first phase of our scheme consists of training a regression model with SVM from labeled data (accounts that have been verified as real or fake). After training, the model is able to output and assign a normalized score ranging from 0 to 1 to each account, which is a rough estimate of the probability each account being fake. Then, the social graph is constructed. From the initial scores obtained from the regression model, we can better characterize the network to produce more precise output, rather than just randomly or uniformly assign a number to each node. A number of iterations of SybilWalk algorithm is then carried out to calculate the final probability score for each node. A higher score means the node is more likely to be a Sybil.
\subsection{Features selection for regression model}
There are some features that we mimic from \cite{intrusion,automatedSybil,junkmail}. We have eliminated some by using entropy \cite{shannon1948} analysis and add some more based on the Facebook specification. The final chosen features are showed in the following table.
\begin{supertabular}{|p{3.5cm}|p{4.5cm}|}
	\hline
	\textbf{Feature} & \textbf{Justification} \\ 
	\hline
	\hline
	How long an account has been active & Fake accounts can be mass produced and are usually only active for a short time. \\ 
	\hline
	Number of friends a user has & Real accounts are expected to make more friends. \\ 
	\hline
	 Number of groups a user has joined & Fake accounts usually join a lot of groups to post spam. \\ 
	\hline
	 Number of posts a user has made & Fake accounts normally do not bother with writing own posts. \\ 
	\hline
	 Number of posts on a user's wall & Fake accounts normally do not bother with posts on walls \\ 
	\hline
	 Number of posts a user has been tagged in & Real accounts have much higher chance to be tagged in other users' posts. \\ 
	\hline
	 Number of times a user has reacted to a post & Fake accounts, especially controlled by a bot are expected to react to a post much more often than real accounts. \\ 
	\hline
	 Number of comments a user has made & Spam messages can also take the form of comments, so fake accounts are likely to make greater number of comments than real accounts. \\ 
	\hline
	 Number of likes all posts of a user has received & Spam messages posted by fake accounts are unlikely to be liked by users. \\ 
	\hline
	 Number of comments on every posts of a user & Spam messages posted by fake accounts are unlikely to receive comments by users. \\ 
	\hline
	 Number of times a user's posts have been shared & Spam messages posted by fake accounts are unlikely to be shared by users. \\ 
	\hline
	 Number of tags on a user's posts (other users and pages alike). & Real accounts use tags much more frequently. \\ 
	\hline
	 Number of users that a user has tagged in his or her posts & Real accounts use tags much more frequently.  \\ 
	\hline
	 Number of pages that a user has tagged in his or her posts & Fake accounts may tag more pages to popularize them. \\ 
	\hline
	 Number of posts that a user has shared & Naturally fake accounts has a much greater share count. \\ 
	\hline
	Number of users that a user has tagged in his or her comments & Real users have real friends, therefore they tag and are tagged much more frequently. \\ 
	\hline
	Number of times a user has been tagged in other users' comments & Real users have real friends, therefore they tag and are tagged much more frequently. \\ 
	\hline
	Number of pages that a user has tagged in his or her comments & Again, fake accounts may tag more pages to popularize them. \\ 
	\hline
\end{supertabular}
%

Extracting and selecting meaningful features from user identities and activities is a crucial but difficult and time consuming task. A lot more features may be taken into consideration, however they may be difficult to extract or completely non-present due to privacy settings. 
\subsection{Building the social graph}
\begin{center}
	\begin{figure}[h!]
		\begin{tikzpicture}
		
		\vertex[](p1) at (-0.7, 0.3) {};
		\vertex[](p2) at (-1.2, 0.5) {};
		\vertex[](p3) at (-1.5, 0.8) {};
		\vertex[](p4) at (-1.4,-0.5) {};
		\vertex[](p5) at (-2.5,0.3) {};
		
		\vertex[](p6) at (-2,-2) {};
		\vertex[](p7) at (-2.6,-4) {};
		\vertex[](p8) at (0,-3) {};
		\vertex[](p9) at (0,-2) {};
		\vertex[](p10) at (-2.4,-3.5) {};
		\vertex[](p11) at (-2.45,-2.5) {};
		\vertex[](p12) at (-1,-3) {};
		\vertex[](p13) at (-0.5,-3.5) {};
		
		\draw [] (p1) to (p2); \draw [] (p1) to (p4); \draw [] (p1) to (p5); \draw [] (p4) to (p6); \draw [] (p4) to (p3); \draw [] (p4) to (p5);
		\draw [] (p3) to (p6);
		\draw [] (p2) to (p4);\draw [] (p2) to (p3); \draw [] (p5) to (p3);
		\draw [] (p6) to (p10);\draw [] (p11) to (p10); \draw [] (p11) to (p6); \draw [] (p9) to (p6); \draw [] (p7) to (p11);
		\draw [] (p9) to (p13);\draw [] (p9) to (p10);\draw [] (p9) to (p12);
		\draw [] (p10) to (p12);\draw [] (p9) to (p11); \draw [] (p12) to (p13); \draw [] (p8) to (p12); \draw [] (p8) to (p9); \draw [] (p7) to (p10);
		
		\vertex[fill = black](p14) at (2,0.5) {};
		\vertex[fill = black](p15) at (2.5,0) {};
		\vertex[fill = black](p16) at (1.3,-0.1) {};
		\vertex[fill = black](p17) at (1.5,-0.8) {};
		
		\draw [] (p14) to (p15); \draw [] (p16) to (p17);  \draw [] (p14) to (p17);
		\draw [] (p16) to (p15); \draw [] (p16) to (p1);
		
		\vertex[fill = black](p18) at (2.1,-3) {};
		\vertex[fill = black](p19) at (1,-3.5) {};
		\vertex[fill = black](p20) at (1.3,-3.8) {};
		\vertex[fill = black](p21) at (1.1,-2.7) {};
		
		\draw [] (p18) to (p17); \draw [] (p18) to (p20);\draw [] (p18) to (p21); \draw [] (p19) to (p21); \draw [] (p19) to (p20); \draw [] (p19) to (p15);
		\draw [] (p19) to (p9);
		
		\node[draw] at (-1.7, -5){Benign Region};
		\node[draw] at (1.8, -5){Sybil Region};
		\node[fill = white] at (-1.6, -3.6){$w_{uv}$};
		
		\node[draw, ellipse] at (0.5, 0.7) {Attack Edge};
		
		\vertex[] (lb) at (-3.5, -1){};
		\vertex[fill = black] (ls) at (4, -1){};
		\draw [dashed] (p1) to (lb);\draw [dashed] (p2) to (lb);\draw [dashed] (p3) to (lb);\draw [dashed] (p4) to (lb);\draw [dashed] (p5) to (lb);\draw [dashed] (p6) to (lb);\draw [dashed] (p7) to (lb);\draw [dashed] (p8) to (lb);\draw [dashed] (p9) to (lb);\draw [dashed] (p10) to (lb);\draw [dashed] (p11) to (lb);\draw [dashed] (p12) to (lb);\draw [dashed] (p13) to (lb);
		
		\draw [dashed] (p14) to (ls);\draw [dashed] (p15) to (ls);\draw [dashed] (p16) to (ls);\draw [dashed] (p17) to (ls);\draw [dashed] (p18) to (ls);\draw [dashed] (p19) to (ls);\draw [dashed] (p20) to (ls);\draw [dashed] (p21) to (ls);
		\node[draw] at (-3.8, -0.3){$l_b$};
		\node[draw] at (4.2, -0.3){$l_s$};
		\end{tikzpicture}
		\caption{\textit{Label-augmented social network}}
		\label{fig:auggraph}
	\end{figure}
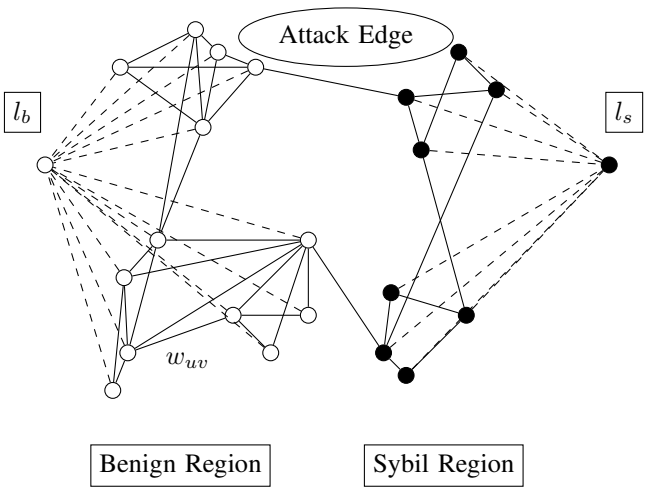
\end{center}
In order to leverage random walks, a model called label-augmented social network \cite{JWZ17} was build. The model consists of the usual social graph, which can be divided into benign region (the subgraph induced by benign nodes) and Sybil region (the subgraph induced by Sybil nodes). Then, two nodes are added to represent each label. We denote by $l_b$ the benign label node and $l_s$ the Sybil label node. $l_b$ and $l_s$ are connected to every nodes of their corresponding label (see Fig. \ref{fig:auggraph}). Each edge is given a weight $w_{uv}$ which is the number of mutual friends between the two nodes, normalized by the maximum number of mutual friends. As for $l_b$ and $l_s$, every edges connected to them are assigned weight $1$. Learning edge weights to better characterize structural relation between nodes is an interesting future direction as well.
\subsection{Calculating and assigning probability score to each node using random walks}
\begin{figure}[h!]
	\begin{center}
		\hspace*{-0.5cm}
		\begin{tikzpicture}
		\node[align=center] at (0, 0){Facebook\\accounts};
		\node[draw] at (2.5, 0){SVM};
		\node[draw, align=center] at (5.2, 0){Building the\\network graph};
		\node[draw] at (8, 0){SybilWalk};
		
		\draw[->] (0.8, 0) -- (1.7, 0);
		\node[align=center] at (1.3, -0.5){Features};
		\draw[->] (3.2, 0) -- (4, 0);
		\node[align=center] at (3.5, -0.5){Initial\\estimates};
		\draw[->] (6.5, 0) -- (7, 0);
		\end{tikzpicture}
	\end{center}
	\caption{\em Model of computation}
	\label{fig:scheme}
\end{figure}
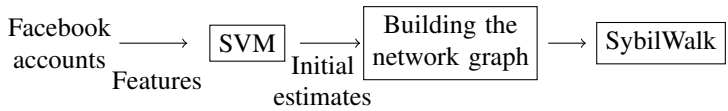

Intuitively, if a node is structurally close to known Sybils, it must have a high probability of being a Sybil itself. We can theoretically perform any number of random walks starting from a node $u$. At each step, the random walks picks a neighbor $v$ of $u$ with probability $\displaystyle\frac{w_{uv}}{\sum_{t\in adj(u)}w_{ut}}$.The probability of $u$ being a Sybil, is the probability of this random walk reaching $l_s$ before $l_b$. This makes efficient use of both social graph structure as well as the ground truth (the known real and fake nodes). However, in implementation, performing so many random walks is impractical because the number of random walks should be sufficiently large to approximate the probability score with high confident, and there is no real way to know how many is "sufficiently large" for a particular graph.\\ 
In \cite{JWZ17}, the authors addressed this problem with an algorithm to compute the score probability of each node via a weighted combination of neighboring nodes. Suppose $u$ has $k$ neighboring nodes $v_1, v_2, ..., v_k$ with probability score $p_1, p_2, ..., p_k$ respectively. If from $u$, the random walk reaches $v_i$ with probability $p_{uv_i}$, then it reaches $l_b$ via $u_i$ with probability $p_{uv_i}p_i$. By law of total probability, we can calculate the probability score for $u$ by
\[p = \sum_{i=1}^{k} p_{uv_i}p_i \]
where $p_{uv_i} = \displaystyle\frac{w_{uv_i}}{\sum_{t\in adj(u)}w_{ut}}$ as mentioned before is the probability a random walk chooses $v_i$ as the next step from $u$. This is the general idea behind the SybilWalk algorithm.
\begin{algorithm}
	\textbf{Input   \hspace{0.05cm}  :} Label-augmented, $\epsilon$ and $T$\\
	\textbf{Output:} $p_u$ for every $u$
	\caption{SybilWalk}
	\begin{algorithmic}[1]
		\State Initialize $p_u^{(0)}$ for every $u$
		\algstore{1}
	\end{algorithmic}
\end{algorithm}
\begin{algorithm}
	\begin{algorithmic}[1]
		\algrestore{1}
		\State Initialize $p_{lb}^{(0)} = 0$
		\State Initialize $p_{ls}^{(0)} = 1$
		\State Initialize $t = 1$
		\While $\sum_{u}(p_u^{(t)}-p_u^{(t-1)})^2 \ge \epsilon$ \&\& $t < T$
		\For {$u$ in $V$}
		\State $p_u^{(t)} = \sum_{v\in adj(u)}\displaystyle\frac{w_{uv}}{\sum_{t\in adj(u)}w_{uv}}p_u^{(t-1)}$
		\EndFor
		\State $t = t + 1$
		\EndWhile
	\end{algorithmic}
\end{algorithm}
The convergence of SybilWalk algorithm is only relative. Therefore, it is important to have a good initial guess. This is why we use SVM to obtain the initial probability scores for each accounts and refine them using random walk. Our model of computation can be summarized in diagram \ref{fig:scheme}.
\section{Implementation and evaluation}\label{result}
\subsection{Acquiring and labeling data}
The need for fake account analysis arose when we were looking into Facebook rumors and communication crises. A rumor breaks out when there are a considerable number of posts circulating about the same subject, attracting many people to comment, like and share. However there may be malicious seeders who want to direct the rumor's spreading to their liking. They will almost always use fake accounts for this purpose, and it's crucial to identify these accounts from genuine ones. We ran a crawler to collect all posts on Facebook regarding some controversial incidents in 2017 - 2018 in Vietnam. Then, we extracted all the accounts (around ten thousands) which participated in these posts. Following that, we proceeded to label the accounts to train and validate our detection scheme. The information of those accounts were acquired using Facebook's User Profile API \cite{API}.
%
\subsection{Evaluation}
We ran our model of computation with 5-fold cross validation. The best result is given in Table \ref{tab:resultcombine}.
\begin{center}
	\begin{table}[H]
		\centering
		\begin{tabular}[c]{|c|c|c|} 
			\hline
			& Fake accounts & Real accounts\\
			\hline
			Precision & 0.9 & 0.96 \\
			\hline
			Recall & 0.85 & 0.97 \\
			\hline
			F1 & 0.87 & 0.96 \\
			\hline
		\end{tabular}
		\caption{\em Precision, Recall and F1 score for the combined model}
		\label{tab:resultcombine}
	\end{table}
\end{center}
The model converged after only over 50 iterations. Compare this with the result when we use only SVM instead of the two-phase scheme in Table \ref{tab:resultSVM}
\begin{center}
	\begin{table}[H]
		\centering\begin{tabular}[c]{|c|c|c|} 
			\hline
			& Fake accounts & Real accounts\\
			\hline
			Precision & 0.8 & 0.92 \\
			\hline
			Recall & 0.73 & 0.95 \\
			\hline
			F1 & 0.76 & 0.94 \\
			\hline
		\end{tabular}
		\caption{\em Precision, Recall and F1 score for SVM detection}
		\label{tab:resultSVM}
	\end{table}
\end{center}
It is obvious that a better performance has been achieved.
\section{Conclusion and perspectives}\label{conclusion}
In this work, we have proposed a ranking scheme for the detection of fake Facebook user accounts which incorporates both feature-based approaches and graph-based approaches to overcome their respective limits. Normalized SVM output first give a rough estimate on the probability score, providing a better initial guess for the SybilWalk algorithm. The computational cost is moderate and can be scaled and deployed to handle large data sets. For future work, there are a few aspects to improve, for example
\begin{itemize}
	\item Learning edge weights to better represent the relationship between nodes.
	\item Evaluate the impacts of features chosen to characterize fake accounts.
	\item Real time detection for application.
\end{itemize}
The source code and the data set can be found at \url{https://github.com/nhisnow1996/Facebook-Fake-Account-Detection}.
\section*{Acknowledgment}
The first author also has receive the support from Institute of Mathematics, Vietnam Academy of Science and Technology, Year 2019. 	This work is also supported by iCOMM Media \(\&\) Tech, Jsc. We would like to thank the iCOMM RnD team for supported resources and text data that we used during training and experiments our model.

\end{document}